\documentclass[conference]{IEEEtran}

\usepackage{graphicx}
\usepackage{epstopdf}
\usepackage{subfig}
\usepackage[cmex10]{amsmath}
\usepackage{algorithmic}
\usepackage{algorithm}
\usepackage{amssymb}
\usepackage{cite}
\usepackage{amsthm} 
\usepackage{xcolor}

\usepackage{multirow}
\usepackage{url}
\usepackage{bm}
\usepackage{mathtools}
\usepackage{tabularx}

\usepackage{url}

\newtheorem{definition}{\textbf{Definition}}

\newcommand{\RN}[1]{%
  \textup{\uppercase\expandafter{\romannumeral#1}}%
}
\newcommand{\q}[1]{``#1''}
\begin{document}
\setlength{\intextsep}{-5pt}

\title{PolarDenseNet: A Deep Learning Model for CSI Feedback in MIMO Systems}

\author{\IEEEauthorblockN{Pranav Madadi\IEEEauthorrefmark{1},
		Jeongho Jeon\IEEEauthorrefmark{2}, Joonyoung Cho,
		Caleb Lo, Juho Lee, Jianzhong Zhang}
	\IEEEauthorblockA{Standards and Mobility Innovation Lab, Samsung Research America, USA\\
		Email: \IEEEauthorrefmark{1}p.madadi@samsung.com,
		\IEEEauthorrefmark{2}jeongho.j@samsung.com}}
\maketitle

\maketitle	




\begin{abstract}
	In multiple-input multiple-output (MIMO) systems, the high-resolution channel information (CSI) is required at the base station (BS) to ensure optimal performance, especially in the case of multi-user MIMO (MU-MIMO) systems. In the absence of channel reciprocity in frequency division duplex (FDD)  systems, the user needs to send the CSI to the BS. Often the large overhead associated with this CSI feedback in FDD systems becomes the bottleneck in improving the system performance. In this paper, we propose an AI-based CSI feedback based on an auto-encoder architecture that encodes the CSI at UE into a low-dimensional latent space and decodes it back at the BS by effectively reducing the feedback overhead while minimizing the loss during recovery. Our simulation results show that the AI-based proposed architecture outperforms the state-of-the-art high-resolution linear combination codebook using the DFT basis adopted in the 5G New Radio (NR) system.
\end{abstract}

\section{Introduction}

The massive multiple-input multiple-output (MIMO) is widely regarded as one of the key technologies in the fifth-generation wireless communication system. With a larger antenna array, such a system can boost both spectrum and energy efficiency, and further support higher-order multi-user (MU)-MIMO transmission to maximize the system performance.

One of the key components of a MIMO transmission is the channel state information (CSI) acquisition at the base station (BS). In frequency division duplex (FDD) systems, the CSI is first estimated at the user equipment (UE) using the CSI reference signal (CSI-RS) transmitted from BS and is then fed back to the BS. The quality of this CSI feedback affects the efficacy of downlink (DL) multi-user (MU) precoder selection at the BS, i.e., directing beams for a selected set of co-scheduled UEs with minimal interference between MU layers. Thus, for MU-MIMO with a high number of antenna ports, the availability of accurate CSI is necessary to guarantee high system performance.

The UE can either feedback the entire CSI matrix referred to in this paper as full channel state information (F-CSI) which is very expensive in terms of uplink (UL) feedback overhead or can pre-process the CSI and send relevant information such as the channel eigenvectors, i.e., dominant beam directions that can directly be used for the design of DL precoder.

The conventional CSI feedback framework adopted in 3rd Generation Partnership Project (3GPP) standards pre-processes the CSI. In particular, a CSI codebook is designed to express channel eigenvectors, with a reduced amount of feedback.

Note that in the case where the channel has multiple dominant eigenvectors i.e, higher rank channel, each eigenvector is fed back independently. Thus, despite the promising performance of the NR high-resolution CSI feedback, the overhead can still be expensive, especially in higher-rank channels.  Therefore, it is desirable to have an enhanced framework to reduce the feedback overhead without hampering the system performance.

A huge body of work also focuses on UE sending the entire CSI, i.e., F-CSI to the BS by reducing feedback overhead by either (a) traditional compressive sensing (CS) methods, and/or (b) deep learning methods, where the CSI is transformed and represented in a sparse domain. The traditional compressed sensing (CS) (LASSO l-1 solver \cite{daubechies2004iterative}, AMP\cite{donoho2009message}, TVAL3\cite{li2009user}) works poorly since the channel matrix may not be sparse enough to achieve higher compression. Furthermore, it is difficult to apply them in the actual communication system due to high computational complexity and time consumption.

In recent years, AI technology has gradually matured and achieved great success in various fields and has also been adopted to solve challenging communication problems, as they have the capability to deal with non-linearity in the system which is usually difficult to mathematically formulate and analytically solve. Furthermore, with deep learning methods achieving great success in the image compression task, it motivated researchers to propose a neural network-based CSI compression to improve the accuracy of full channel state information (F-CSI) feedback with reduced overhead that alleviates the problems of previous CS methods.

An architecture based on autoencoder, i.e., CsiNet \cite{wen2018deep} was proposed to compress CSI and reconstruct it, providing performance gains in comparison with CS-based methods. CsiNet has two components, (a) \textit{encoder} at UE that transforms the CSI into a compressed latent space leveraging the channel structure learned through data training, and (b) \textit{decoder} at BS, that recovers the original channel from the compressed representation. This work inspired a lot of research, such as CsiNet-LSTM \cite{wang2018deep}, where, to better extract the correlation of CSI in the time domain, long short term memory (LSTM) was combined with CsiNet. Multi-user cooperative feedback was considered in \cite{guo2020dl}, and some end-to-end models are designed combining CSI feedback with channel estimation or beamforming \cite{guo2020deep}, \cite{chen2020deep}. To improve the accuracy of CSI feedback, a multi-resolution convolution block was designed in the CRNet \cite{lu2020multi}. 

 
 A novel architecture called ACRNet \cite{lu2021aggregated} was proposed to further improve the state-of-the-art performance with network aggregation \cite{xie2017aggregated} i.e., a standard convolution block is split into parallel groups and with a learnable activation function.

The above works have achieved great performance in the F-CSI feedback mechanism where the channel is transformed into another space, such as angular-delay domain\cite{wang2018deep}, to leverage the sparsity. However, this is different from the widely adopted conventional CSI feedback scheme in 3GPP protocols that mainly focuses on the feedback of precoder channel data as discussed before. For practical purposes, it is very much desirable to minimize the CSI feedback overhead, hence the codebook-based precoder channel feedback is largely implemented. Therefore, inspired by the successful research of deep learning in F-CSI feedback, we leverage the dual-polarization of antennas, which is very common for commercial cellular BS deployment, and propose a novel AI-based model PolarDenseNet to further enhance performance compared with the conventional feedback scheme.

The paper is organized as follows. In section \RN{2}, we define the precise problem statement and summarize the 5G NR codebook-based CSI feedback schemes in section-\RN{3} that will be the baseline to compare the performance of our proposed AI model. We then introduce the architecture of our proposed model in section \RN{4}. In section \RN{5}, we present our simulation setup with results and conclude in section \RN{6}.

\section{System Model}

In this paper, we consider a massive MIMO system with $N_t >>1$ transmit antennas at a base station (BS) and $N_r$ receive antennas at a user equipment (UE). We assume a single-panel bipolar antenna array (i.e., number of polarizations = 2) in a rectangular shape, where $N_1$  and $N_2$  antenna ports are arranged in horizontal and vertical directions respectively, as depicted in Fig. 1. In this figure, each antenna element is logically mapped onto a single antenna port. In general, one antenna port may correspond to multiple antenna elements (physical antennas) combined via virtualization. The total number of antenna ports at the BS is $N_t = 2 N_1 N_2$. In 5G NR, the maximum number of supported antenna ports is 32 and a variety of different antenna port layouts are supported \cite{3GPP38.214}.

\begin{figure}[!htpb]
	\centering
	\includegraphics[width=2in]{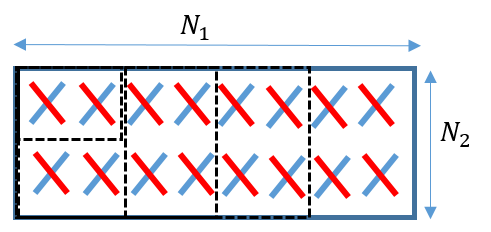}
	\caption{An illustration of dual polarized antenna array at BS for various possible configurations of $(N_1,N_2)$.}
	\label{fig_sim}
\end{figure}

\vspace{0.5cm}
In this paper, we assume an orthogonal frequency division multiplexing (OFDM) system operating over $N_c$ subcarriers. The received signal at the $n^{th}$ subcarrier is

\begin{equation}
	\label{basic}
	y_n = h_n^H v_n x_n +z_n,
\end{equation}
where $ y_n \in \mathbb{C}^{N_r \times 1}$, $h_n \in \mathbb{C}^{N_t \times N_r}$, $v_n \in \mathbb{C}^{N_t \times N_s}$, $x_n \in \mathbb{C}^{N_s \times 1}$, $z_n \in \mathbb{C}^{N_r \times 1}$ are the received symbol, channel matrix, precoding matrix, transmitted symbol and AWGN noise, respectively. Note that $N_s$ represents the total number of modulated symbols transmitted at $n^{th}$ subcarrier, that can take values in the range $[1, \mbox{min}(N_r,N_t)]$. The channel matrix stacked in frequency domain is then referred to as F-CSI, $H = \left[h_1 \dots h_{N_c} \right] \in \mathbb{C}^{N_c \times N_t \times N_r}$.

The conventional 3GPP scheme pre-processes the F-CSI by performing the eigenvalue decomposition of the channel since the dominant beam direction for a particular frequency subband corresponds to the dominant eigenvector of the channel. Thus, with the bandwidth split into various subbands, the UE sends these dominant eigenvectors as the CSI feedback. Now, let us formally define the precoder channel matrix as follows.

\begin{definition}
	In MIMO OFDM systems with $N$ transmit antennas and with bandwidth split into $K$ subbands each containing a certain number of resource blocks, the precoder channel matrix $H_{N,K}$, is constructed by stacking $K$ dominant eigenvectors of the precoder of size $N \times 1$, each corresponding to a specific subband.  
\end{definition}

For simplicity, in the rest of the paper, we refer to the number of transmit antennas as $N$ instead of $N_t$.

The feedback overhead for the entire precoder channel matrix $H_{N,K}$, i.e., the number of real-valued feedback parameters $Z = 2 \times N \times K$,  become too large to be reported over the available uplink (UL) channels. Therefore, the precoder channel matrix $H_{N,K}$  needs to be “compressed” before being fed back. Thus, the problem statement is to propose a scheme that can be practically implemented and reduce the CSI feedback overhead, i.e., a scheme that can be an alternate means to the existing conventional feedback framework.

In this paper, we propose to use an AI-based auto-encoder to compress the precoder channel matrix $H_{N,K}$ and then perform a system-level simulation to compare its performance with the conventional 3GPP schemes, details of which are mentioned in the following section.

\section{Baseline: 3GPP codebook based feedback}

In this section, we briefly describe the 5G NR CSI codebook design and feedback framework. In Rel 15 type-\RN{2} codebook scheme \cite{3GPP38.214}, the CSI compression was only in the spatial domain (SD) where spatial Discrete Fourier transform (DFT) basis set is generated at UE based on the antenna configuration and oversampling factors. After that, $L$ orthogonal DFT beams are selected which are common to both antenna polarization and across sub-bands. Dominant eigenvectors per subband are represented as a linear combination (LC) of these selected $L$ DFT beams.

Let $\textbf{B} = [b_0, \dots b_{L-1}]$ denote the beam matrix whose columns are the $L$ selected beams. Then, the linear combination coefficients are given by

\begin{equation}
	\label{SD}
	W_2 = inv(W_1)H_{N,K}=  W_1^H H_{N,K},
\end{equation}
where $W_1 = \begin{bmatrix}
	\textbf{B} & 0 \\
	0 & \textbf{B}
\end{bmatrix}$ have orthogonal column vectors therefore, Hermitian of the matrix is also its inverse. 

\begin{figure}[!tbhp]
	\centering
	\includegraphics[width=3.5in]{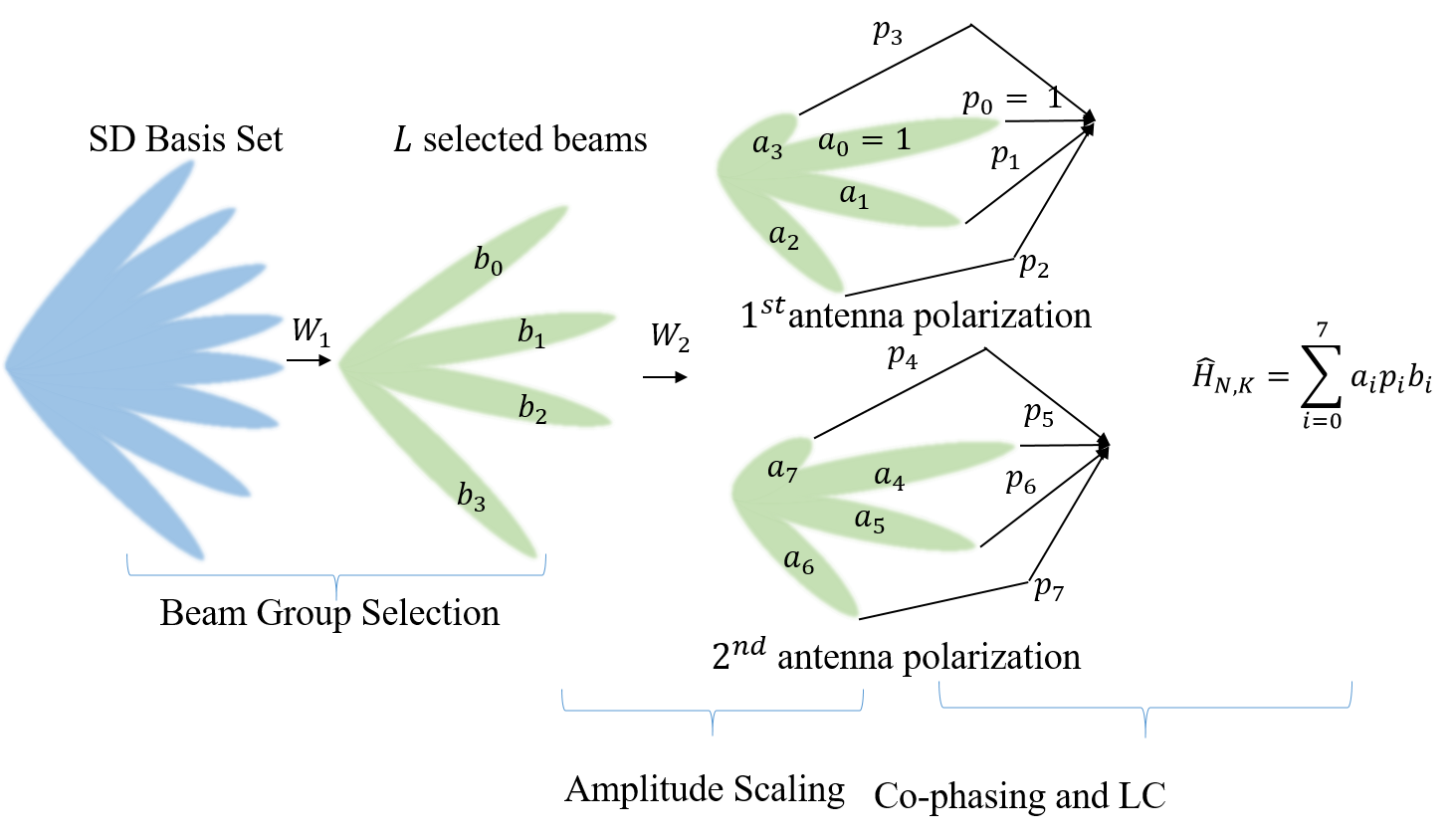}
	\caption{Illustration of Rel-15 type-\RN{2} CSI codebook based feedback scheme.}
	\label{rel15_plot}
\end{figure}

The Rel-15 codebook can be viewed as a dual-stage CSI codebook, where the reconstructed precoder channel matrix is given by
\begin{equation}
	\label{rel15}
	\hat{H}_{N,K} = W_1 W_2,
\end{equation}
where $W_1$ comprises of orthogonal beam selection and $W_2$ comprises of subband coefficient amplitude and phase that is different for each antenna polarization. In Eq. \eqref{rel15}, normalization was omitted to simplify the description. This dual-stage CSI codebook is illustrated pictorially in Fig. \ref{rel15_plot}.

In a later release, Rel-16 type-\RN{2} CSI feedback \cite{3GPP38.214}, $H_{N,K}$ is compressed jointly in both spatial and frequency domain (FD).  In particular, the correlation in the rows of $W_2$ is exploited. To that extent, the compression in the FD domain is performed by LC of $M$ FD DFT basis vectors, i.e. columns of $W_f$ for FD compression as in

\begin{equation}
	C = W_2 W_f,
	\label{coeff}
\end{equation}
where $W_f = \left[f_0 \dots f_{M-1} \right]$. The UE reports SD, FD basis vectors, $W_1$, $W_2$ and $C$ comprises of frequency and spatial domain coefficients. Then, the reconstructed precoder channel matrix for Rel-16 is given as

\begin{equation}
	\hat{H}_{N,K} = W_1CW^H_f,
\end{equation}
where $C$ is the coefficient matrix composed of linear combination coefficients as in Eq.\eqref{coeff}.

In this paper, we show that our proposed architecture outperforms the above described 5G NR high-resolution codebook-based CSI feedback in both metrics of comparison, i.e., NMSE and cosine-similarity defined in later section, with reduced feedback bits.

\section{Proposed Architecture}

\begin{figure}[!htpb]
	\centering
	\includegraphics[width=3in]{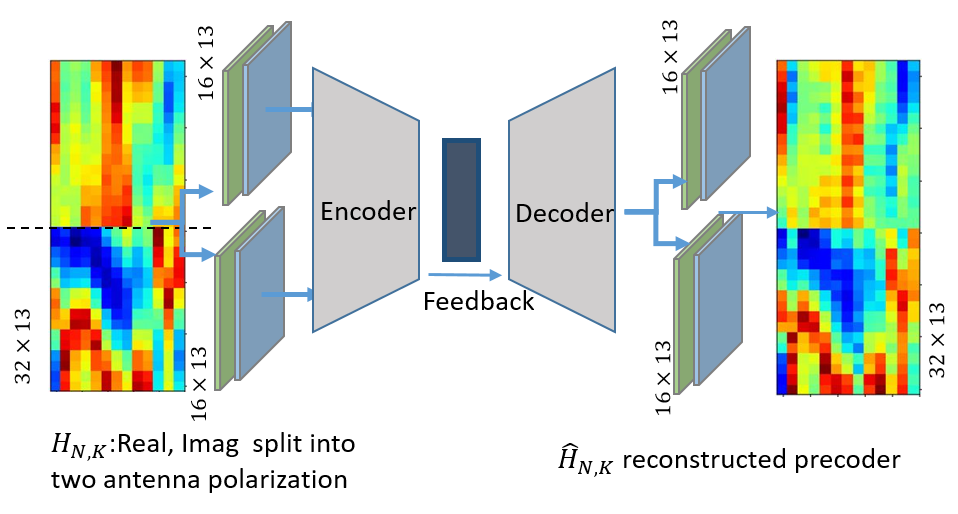}
	\caption{Proposed autoencoder architecture.}
	\label{autoencoder_overall}
\end{figure}
In the field of deep learning, autoencoders have gained a lot of interest in recent times. Autoencoder is an unsupervised artificial neural network that learns how to efficiently compress/encode data and then, learns how to reconstruct the data back from the reduced encoded representation to a representation that is as close to the original input as possible. The autoencoder achieves this compression by exploiting the correlation in the original input data.

Note that the convolutional layer plays a significant role in neural networks especially for computer vision tasks. Since the CSI matrix can be viewed as an image depicting the physical layer channel pattern, most previous work on AI-based CSI feedback such as CsiNet and ACRNet, are based on a convolutional neural network (CNN) \cite{wen2018deep}. 

Thus, in this paper, we propose to use an auto-encoder framework based on CNNs for CSI feedback named \textit{PolarDenseNet} as illustrated in Fig.\ref{autoencoder_overall}, which further improves feedback compression and recovery performance over existing schemes by exploiting dual-polarized antenna structure and various recent neural network architectures and techniques. 
\begin{figure*}[!htpb]
	\centering
	\includegraphics[width=5.5in]{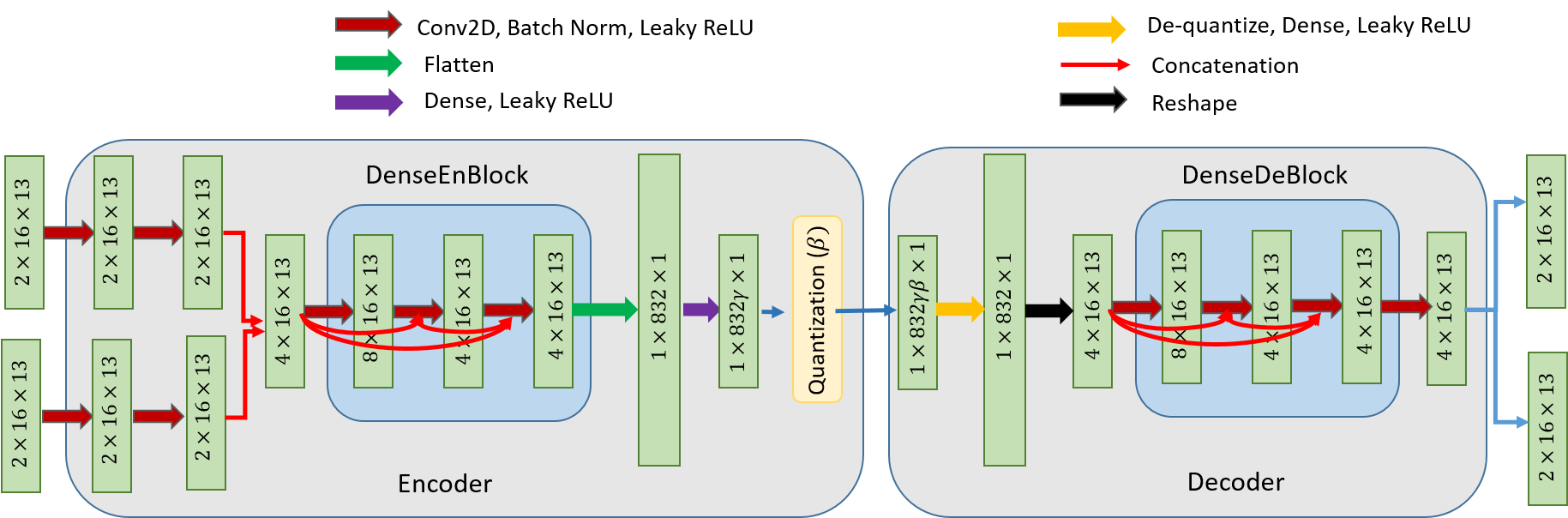}
	\caption{Encoder and Decoder architecture for the proposed PolarDenseNet. The input shape (number of input channels, height, width) is mentioned on the blocks with thick arrows representing various composite NN layers and thin arrows representing simple concatenation, split, or fowarding operations.}
	\label{autoencoder}
\end{figure*}

Autoencoder has two main components, an encoder that takes the precoder channel matrix as input and transforms it into a compressed codeword (i.e., a bitstream) which is sent by UE as CSI feedback through the UL feedback channel. It is assumed that the compressed codeword is received by BS without an error owing to the error-correcting codes and re-transmission schemes. Then, at BS a decoder takes the compressed codeword as input and reconstructs the precoder channel matrix. 

Let $\theta=\{\theta_{en},\theta_{de}\}$ represent the set of trainable parameters of the machine learning model, then the reconstructed precoder channel $\hat{H}_{N,K}$ at the  BS can be represented as
\begin{equation}
	\hat{H}_{N,K}=f_{de}(Q(f_{en}(H_{N,K};\theta_{en}) );\theta_{de}),
	\label{reconstruct}
\end{equation}
where $f_{en} (.), f_{de}(.)$ represents the encoder and decoder respectively and $Q(.)$ represents the quantization operation. For training, we perform end-to-end learning for all the kernels and bias values of the encoder and decoder jointly.  The goal is to minimize the mean square error (MSE) given as

\begin{equation}
	\label{lossfuntion}
	MSE = \frac{1}{S} \sum_{i=1}^{S} ||\hat{H}^i_{N,K} - H^i_{N,K}||_F^2,
\end{equation}
where $S$ represents the total number of samples in the training set. The final feedback overhead, i.e., the number of feedback bits, is decided by the compression ratio, $\gamma$, and quantization bits, $\beta$, i.e. the number of bits used to represent a real number, as
\begin{equation}
	N_{bits} = Z \times \gamma \times \beta,
\end{equation}
where $Z$ is the original dimension of $H_{N,K}$, i.e., $2 \times N \times K$.

To evaluate the reconstruction performance we adopt the widely used metric in the DL-based CSI feedback, the
normalized MSE (NMSE) given by

\begin{equation}
	\eta = \mathbb{E}\left[ \frac{||\hat{H}_{N,K} - H_{N,K}||_F^2}{||H_{N,K}||_F^2}   \right].
\end{equation}

We also consider the cosine-similarity as another performance metric, given by

\begin{equation}
	\rho = \mathbb{E} \left[ \frac{1}{K} \sum_{k=1}^{K} \frac{|\hat{h}_k^H h_k|}{||\hat{h}_k||_2|| h_k||_2} \right],
	\label{cosinesim}
\end{equation}
where $\hat{h}_k, h_k$ are the column vectors of the reconstructed and original precoder channel matrix, respectively. The range of values for $\rho$ is $[0,1]$, with values closer to $1$ implying that the reconstructed vector is similar to original precoder vector.

The UE first compresses the estimated CSI and then quantizes the compressed codeword to further reduce the feedback overhead. Once the BS obtains the feedback bits, the BS decompresses the received feedback bitstream into the original CSI form and utilizes it for MIMO beamforming precoder selection.  Hence, the key point of the architecture is the design of neural networks (NNs) for stronger precoder channel matrix compression and recovery ability. To this end, we demonstrate our proposed model in detail in the following subsection.

\subsection{Details about the PolarDenseNet NNs}

Motivated by the architecture of CSINet and ACRNet autoencoders that are shown to have great performance with F-CSI feedback on COST-2100 model, we propose our architecture as illustrated in Fig.\ref{autoencoder}.

The input to the encoder is the precoder channel matrix $H_{N,K} \in R^{2 \times N \times K}$, where the value 2 represents the real and imaginary parts of the matrix. Given the dual-polarization of the antennas, we expect a high correlation within a given polarization, but less correlation across the channels between two different polarization. An abrupt transition between the upper half and lower half channel matrix is observed as illustrated in Fig.\ref{autoencoder_overall}, because the upper and lower halves of the channel matrix represent the first and second antenna polarization, respectively.

Thus, we extract features separately by splitting the input matrix into two halves and sending them to two different convolution paths, i.e., convolution layers. In each path, there are two convolutional layers with different kernel sizes. The first one has a kernel size of $8 \times 1$ and the second one $1 \times 8$. The kernel sizes are designed to capture the spatial and frequency domain correlation, separately. Also, it is proven to have less complexity in terms of parameters to have two convolutional layers with $1 \times n$, $n \times 1$ kernel sizes than to have one layer with $n \times n$ kernel \cite{szegedy2016rethinking}.

Later, we concatenate the features from both convolution paths. The reason for this joint processing is that the underlying physics governing the channel states between the two polarizations are the same, i.e., one antenna element is stacked on top of another antenna element having the opposite polarization. Therefore, there exists a correlation between the first half and the second half channels, which will be extracted using the kernels after concatenation. 

We further concatenate the features to enable maximum feature retain-ability and have a dense encoder block (DenseEnBlock) that is inspired from the dense convolution networks \cite{huang2017densely}, which connects each layer to every other layer in a feed-forward fashion. These networks have several compelling advantages: they alleviate the vanishing gradient problem, strengthen feature propagation, encourage feature reuse, and substantially reduce the number of parameters. The DenseEnBlock consists of three convolution layers and output feature concatenation. We further define dense decoder blocks (DenseDeBlocks) used in the decoder that has similar architecture. The integration of these blocks is illustrated in Fig. \ref{autoencoder}.

Following a flatten layer at the encoder, the final layer of the encoder is a dense layer that reduces the dimension of the input to the desired length. One can implement pooling to reduce the dimensions, i.e., a form of downsampling, and use transpose convolutions \cite{dumoulin2016guide} later at the decoder to upscale, but the current architecture using dense layers yielded better results in our experiments.

All convolutional layers have zero padding on the perimeter of the input matrix to maintain the dimensions of the matrix and are followed by batch normalization and Leaky rectified linear activation unit (LReLU) activation layer. The batch normalization layer has proved to affect the training of the model, dramatically reducing the number of epochs required. It can also have a regularizing effect, reducing generalization error much like the use of activation regularization.  On other hand,  the ReLU function has rapidly become the default activation function when developing most types of neural networks, given its benefits over hyperbolic tangent activation (Tanh) and sigmoid activation functions \cite{sharma2017activation}.  We consider LReLU to overcome the \q{dying ReLU} problem where there could be dead neurons in the network since ReLU prunes the negative values to zero. The equation for LReLU is given as

\begin{equation}
	LReLU(x) = \begin{cases}
		x  , x \geq 0, \\
		\alpha x, x <0,
	\end{cases}
\end{equation}
where we set the parameter $\alpha$ to be $0.3$ in this paper. The last step at the encoder is quantization, where we use a simple uniform 2-bit quantization.

\section{Simulation Setup and Results}
In this section, we perform numerical simulations to evaluate and analyze the performance of the proposed AI-based model. First, we elaborate on the simulation settings, including channel model and network training details. Then, we compare the performance under different simulation cases between the proposed AI-model and Rel-16 type II codebook which is considered as the baseline.
\subsection{Experimental Settings}
We generated the precoder channel matrix $H_{N,K}$ using the system level simulator, considering a carrier frequency of $4$ GHz with a bandwidth of $10$ MHz. Further, we considered a  sub carrier spacing of $15$ kHz, thus each transmission time index (TTI) corresponds to $1$ ms with 14 OFDM symbols. Note that our model has $624$ sub-carriers and since each resource block (RB) is of $12$ sub-carriers, our model has $52$ RBs. 

We considered the 3GPP standard 3D Urban Macro (UMa) channel model \cite{3GPP38.901} with inter-site distance (ISD) of $500$ m with uniform linear array (ULA) of $N=32$, transmit antennas and $N_r=4$, receive antennas. Finally, we considered a subband size of $4$ RBs, thus we have the total number of subbands $K$ to be $13$.

We generated the channel matrix at every 5 ms, which is also every 5 TTI given the numerology explained above,  along with the reconstructed channel based on Rel-15 and Rel-16 codebook-based CSI feedback in the system level simulator for multiple UEs and for multiple seed values for improving the generalization of the model.

We used the generated channel as input to the PolarDenseNet for end-to-end training of parameters $\theta$. We trained the model on 70,000 training samples, with 10,000 validation samples, and calculate the normalized mean square error for the reconstructed channel for 10,000 testing samples. The training is performed using Keras TensorFlow with an NVIDIA Quadro RTX 6000 GPU.

As for deep learning hyper-parameters, the batch size is 200
and the loss function is MSE with Adam optimizer. We adopt the warm-up aided cosine annealing scheduler introduced in \cite{lu2020multi}, which can be derived as follows

\begin{equation}
	lr = lr_{min} + \frac{1}{2}(lr_{max}-lr_{min})\left(1+cos\left(\frac{1-T_w}{T-T_w}\pi\right)\right),
\end{equation}
where $T_w$ stands for the index of current epoch and $lr$ is the corresponding learning rate. The number of training epochs is set to $400$ and the number of warm up epochs is set to $30$.  The initial learning rate $lr_{max}$ is set to 1e-2 while the minimal learning rate $lr_{min}$ is set to 1e-4.

\subsection{Simulation results}

\begin{figure}
	\centering
	\includegraphics[width=3in]{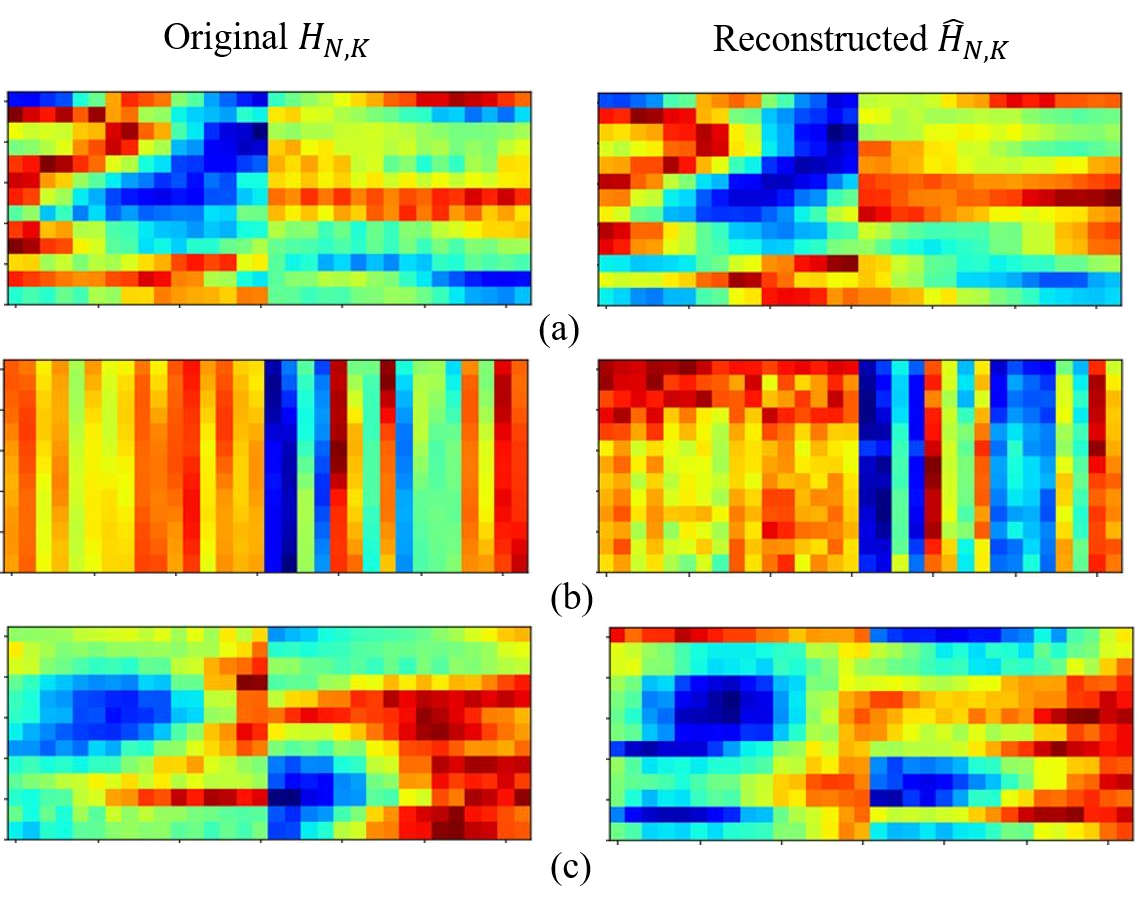}
	\caption{Heat map of the precoder and reconstructed precoder channel matrices, ($H_{N,K}$, $\hat{H}_{N,K}$) for various compression ratios (a) $\gamma = 1/8$, NMSE = -8.85 dB, (b) $\gamma = 1/16$, NMSE = -5.32 dB, (c) $\gamma = 1/20$, NMSE = -4.52 dB.}
	\label{heatmap}
\end{figure}

Note that the model changes in terms of sizes of the neural networks and parameters for different compression ratios, $\gamma$. Thus, we train the PolarDenseNet for various compression ratios, $\gamma = [1/8,1/16,1/20]$. The reconstruction of the precoder channel matrix is better for less compression, i.e., the higher compression ratio of 1/8 compared to that of 1/16. Fig.\ref{heatmap}, illustrates the heat map of the original precoder channel matrix $H_{N,K}$ and the reconstructed channel matrix $\hat{H}_{N,K}$ for various compression ratios. 

\subsection{Comparison with 5G NR schemes}

We compare the performance of CSI feedback of our proposed AI-model with 5G NR codebook-based schemes, i.e, Rel-15/Rel-16 type-\RN{2}, by evaluating both the NMSE ($\eta$) and cosine-similarity ($\rho$). We compare these performance metrics for various configurations of parameters such as the number of SD/FD beams ($L/M$), as supported by 3GPP standards with various compression ratios of the PolarDenseNet as illustrated in Table \RN{1}.

\begin{table}
	\centering
	\label{comparison}
	\caption{NMSE in dB and Cosine-similarity $\rho$ of various schemes}
	\begin{tabular}{ |c |c| c| c| } 
		
		\hline
		Feedback scheme &  NMSE & $\rho$ & Feedback Bits  \\ 
		\hline
		Rel-15, $L=4$  & -3.8 & 0.94 &  351  \\ 
		Rel-15, $L=2$  & -2.1  & 0.89 & 176 \\ 
		\hline
		Rel-16, $L=4, M=3$  & -1.6 & 0.82 & 173\\ 
		Rel-16, $L=2, M=3$  & 0.3 & 0.69 & 58  \\ 
		\hline
		PolarDenseNet, $\gamma=1/8$  & -8.85 & 0.93 & 208 \\ 
		PolarDenseNet, $\gamma=1/16$  & -5.32 & 0.85 & 104 \\ 
		PolarDenseNet, $\gamma=1/20$  & -4.52 & 0.81 & 80 \\ 
		\hline
		
	\end{tabular}
\end{table}

As illustrated by the results, our proposed AI model outperforms the conventional 5G-NR schemes in terms of precoder reconstruction accuracy with less feedback overheard. 

\subsection{Robustness of the model to noise} 

\begin{figure}
	\centering
	\includegraphics[width=3in]{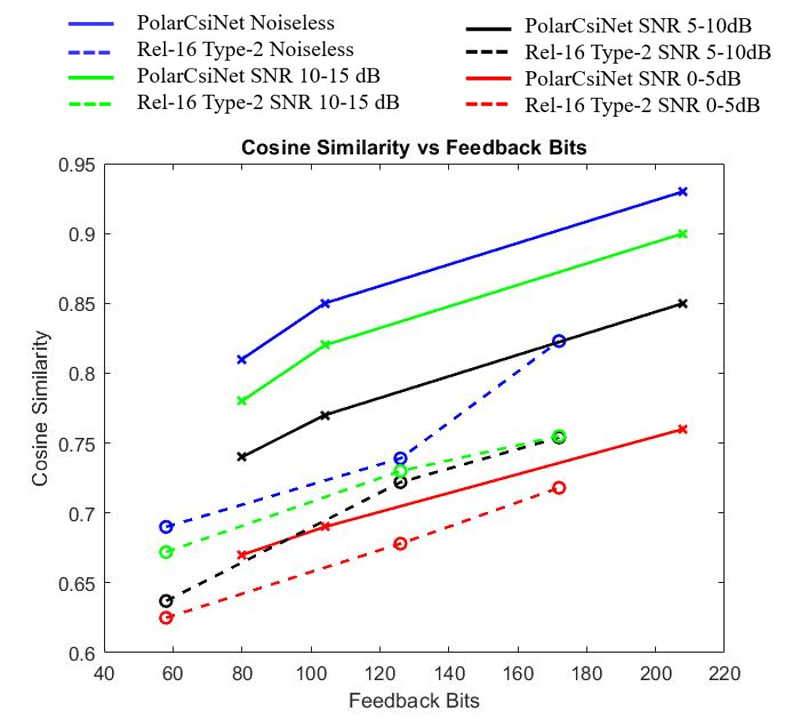}
	\caption{Cosine-similarity results comparing the performance of the PolarDenseNet for various compression ratios $\gamma =[1/8,1/16,1/20]$ with various Rel-16 type-\RN{2} configurations under AWGN noise.}
	\label{cosine}
\end{figure}

To validate the performance of the AI-model under noisy conditions, we simulate CSI feedback by introducing three types of additive Gaussian white noise (AWGN) whose signal-to-noise ratio (SNR) is $0\sim 5$ dB, $5 \sim 10$ dB, and $10 \sim 15$ dB to the precoder channel matrices as input samples. The white noise following the same SNR distribution is applied for both training and testing. Then, Fig. \ref{cosine} illustrates the simulation results comparing the cosine-similarity ($\rho$), where the cosine-similarity of the AI-model is better than Rel-16 type-\RN{2} codebook based schemes for various configurations and SNR ranges.

\section{Conclusion}

In this paper, we propose an AI-based model, PolarDenseNet, for CSI feedback that is inspired from the dense convolutional networks and leverages the dual-polarization of antennas. Numerical results show that it has better reconstruction performance of the precoder channel matrix with reduced feedback overhead than the conventional Rel-15/Rel-16 codebook-based type-\RN{2} CSI feedback. We further demonstrated and confirmed the robustness of the model under AWGN noise. Thus, the proposed framework can successfully replace the state-of-the-art 5G NR codebook-based schemes.

\IEEEpeerreviewmaketitle

\bibliographystyle{IEEEtran}
\bibliography{ref.bib}

\begin{thebibliography}{10}
\providecommand{\url}[1]{#1}
\csname url@samestyle\endcsname
\providecommand{\newblock}{\relax}
\providecommand{\bibinfo}[2]{#2}
\providecommand{\BIBentrySTDinterwordspacing}{\spaceskip=0pt\relax}
\providecommand{\BIBentryALTinterwordstretchfactor}{4}
\providecommand{\BIBentryALTinterwordspacing}{\spaceskip=\fontdimen2\font plus
\BIBentryALTinterwordstretchfactor\fontdimen3\font minus
  \fontdimen4\font\relax}
\providecommand{\BIBforeignlanguage}[2]{{%
\expandafter\ifx\csname l@#1\endcsname\relax
\typeout{** WARNING: IEEEtran.bst: No hyphenation pattern has been}%
\typeout{** loaded for the language `#1'. Using the pattern for}%
\typeout{** the default language instead.}%
\else
\language=\csname l@#1\endcsname
\fi
#2}}
\providecommand{\BIBdecl}{\relax}
\BIBdecl

\bibitem{daubechies2004iterative}
I.~Daubechies, M.~Defrise, and C.~De~Mol, ``An iterative thresholding algorithm
  for linear inverse problems with a sparsity constraint,''
  \emph{Communications on Pure and Applied Mathematics: A Journal Issued by the
  Courant Institute of Mathematical Sciences}, vol.~57, no.~11, pp. 1413--1457,
  2004.

\bibitem{donoho2009message}
D.~L. Donoho, A.~Maleki, and A.~Montanari, ``Message-passing algorithms for
  compressed sensing,'' \emph{Proceedings of the National Academy of Sciences},
  vol. 106, no.~45, pp. 18\,914--18\,919, 2009.

\bibitem{li2009user}
C.~Li, W.~Yin, and Y.~Zhang, ``User’s guide for {TVAL3}: {TV} minimization by
  augmented lagrangian and alternating direction algorithms,'' \emph{CAAM
  report}, vol.~20, no. 46-47, p.~4, 2009.

\bibitem{wen2018deep}
C.-K. Wen, W.-T. Shih, and S.~Jin, ``Deep learning for massive {MIMO} {CSI}
  feedback,'' \emph{IEEE Wireless Communications Letters}, vol.~7, no.~5, pp.
  748--751, 2018.

\bibitem{wang2018deep}
T.~Wang, C.-K. Wen, S.~Jin, and G.~Y. Li, ``Deep learning-based {CSI} feedback
  approach for time-varying massive {MIMO} channels,'' \emph{IEEE Wireless
  Communications Letters}, vol.~8, no.~2, pp. 416--419, 2018.

\bibitem{guo2020dl}
J.~Guo, X.~Yang, C.-K. Wen, S.~Jin, and G.~Y. Li, ``Dl-based {CSI} feedback and
  cooperative recovery in massive {MIMO},'' \emph{arXiv preprint
  arXiv:2003.03303}, 2020.

\bibitem{guo2020deep}
J.~Guo, C.-K. Wen, and S.~Jin, ``Deep learning-based {CSI} feedback for
  beamforming in single-and multi-cell massive {MIMO} systems,'' \emph{IEEE
  Journal on Selected Areas in Communications}, 2020.

\bibitem{chen2020deep}
T.~Chen, J.~Guo, C.-K. Wen, S.~Jin, G.~Y. Li, X.~Wang, and X.~Hou, ``Deep
  learning for joint channel estimation and feedback in massive {MIMO}
  systems,'' \emph{arXiv preprint arXiv:2011.07242}, 2020.

\bibitem{lu2020multi}
Z.~Lu, J.~Wang, and J.~Song, ``Multi-resolution {CSI} feedback with deep
  learning in massive {MIMO} system,'' in \emph{ICC 2020-2020 IEEE
  International Conference on Communications (ICC)}.\hskip 1em plus 0.5em minus
  0.4em\relax IEEE, 2020, pp. 1--6.

\bibitem{lu2021aggregated}
Z.~Lu, H.~He, Z.~Duan, J.~Wang, and J.~Song, ``Aggregated network for massive
  {MIMO} {CSI} feedback,'' \emph{arXiv preprint arXiv:2101.06618}, 2021.

\bibitem{xie2017aggregated}
S.~Xie, R.~Girshick, P.~Doll{\'a}r, Z.~Tu, and K.~He, ``Aggregated residual
  transformations for deep neural networks,'' in \emph{Proceedings of the IEEE
  conference on computer vision and pattern recognition}, 2017, pp. 1492--1500.

\bibitem{3GPP38.214}
3GPP, ``{3GPP} {TS} 38.214 v16.7.0, 3rd {G}eneration {P}artnership {P}roject;
  {T}echnical {S}pecification {G}roup {R}adio {A}ccess {N}etwork; {NR};
  {P}hysical layer procedures for data,'' \emph{Technical Specification}, 2021.

\bibitem{szegedy2016rethinking}
C.~Szegedy, V.~Vanhoucke, S.~Ioffe, J.~Shlens, and Z.~Wojna, ``Rethinking the
  inception architecture for computer vision,'' in \emph{Proceedings of the
  IEEE conference on computer vision and pattern recognition}, 2016, pp.
  2818--2826.

\bibitem{huang2017densely}
G.~Huang, Z.~Liu, L.~Van Der~Maaten, and K.~Q. Weinberger, ``Densely connected
  convolutional networks,'' in \emph{Proceedings of the IEEE conference on
  computer vision and pattern recognition}, 2017, pp. 4700--4708.

\bibitem{dumoulin2016guide}
V.~Dumoulin and F.~Visin, ``A guide to convolution arithmetic for deep
  learning,'' \emph{arXiv preprint arXiv:1603.07285}, 2016.

\bibitem{sharma2017activation}
S.~Sharma and S.~Sharma, ``Activation functions in neural networks,''
  \emph{Towards Data Science}, vol.~6, no.~12, pp. 310--316, 2017.

\bibitem{3GPP38.901}
3GPP, ``{3GPP} {TS} 38.901 v16.1.0, 3rd {G}eneration {P}artnership {P}roject;
  ;study on channel model for frequencies from 0.5 to 100 ghz,''
  \emph{Technical Report}, 2021.

\end{thebibliography}

\end{document}